\documentclass{article} 
\usepackage{iclr2024_conference,times}

\usepackage{amsmath,amsfonts,bm}

\def\eqref#1{equation~\ref{#1}}

\def\1{\bm{1}}

\def\vzero{{\bm{0}}}

\def\vtheta{{\bm{\theta}}}
\def\vomega{{\bm{\omega}}}
\def\vphi{{\bm{\phi}}}
\def\vpsi{{\bm{\psi}}}

\def\va{{\bm{a}}}
\def\vb{{\bm{b}}}

\def\vd{{\bm{d}}}

\def\vk{{\bm{k}}}

\def\vr{{\bm{r}}}

\def\vv{{\bm{v}}}

\def\vx{{\bm{x}}}

\def\mA{{\bm{A}}}

\def\mI{{\bm{I}}}

\DeclareMathAlphabet{\mathsfit}{\encodingdefault}{\sfdefault}{m}{sl}
\SetMathAlphabet{\mathsfit}{bold}{\encodingdefault}{\sfdefault}{bx}{n}

\usepackage{hyperref}
\usepackage{url}
\usepackage{todonotes}
\usepackage{amssymb}
\usepackage{booktabs}
\usepackage{amsmath}
\usepackage{mathtools}
\usepackage{amsthm}
\usepackage {makecell} 
\usepackage{multirow}
\usepackage{algorithm}
\usepackage{algpseudocode}

\theoremstyle{plain}
\newtheorem{theorem}{Theorem}[section]
\newtheorem{Prop}[theorem]{Proposition}
\newtheorem{Lemma}[theorem]{Lemma}

\theoremstyle{definition}
\newtheorem{definition}[theorem]{Definition}

\theoremstyle{remark}

\title{Sliced Denoising: A Physics-Informed Molecular Pre-Training Method}

\author{Yuyan Ni \thanks{Equal contribution. }\ \ \thanks{Work was done while Yuyan Ni was a research intern at AIR.}   \\
Academy of Mathematics and Systems Science\\ 
Chinese Academy of Sciences\\
Beijing, China \\
\texttt{niyuyan17@mails.ucas.ac.cn} \\
\And
Shikun Feng $^\ast$ \\
Institute for AI Industry Research (AIR) \\
Tsinghua University \\
Beijing, China \\
\texttt{fsk21@mails.tsinghua.edu.cn}
 \And
Wei-Ying Ma\\
Institute for AI Industry Research (AIR) \\
Tsinghua University \\
Beijing, China \\
\texttt{maweiying@air.tsinghua.edu.cn}
 \And
Zhi-Ming Ma\\
Academy of Mathematics and Systems Science\\ 
Chinese Academy of Sciences\\
Beijing, China \\
\texttt{mazm@amt.ac.cn}
\And
Yanyan Lan   \thanks{Correspondence to Yanyan Lan $<$lanyanyan@air.tsinghua.edu.cn$>$}\\
Institute for AI Industry Research (AIR) \\
Tsinghua University \\
Beijing, China \\
\texttt{lanyanyan@air.tsinghua.edu.cn}\\
}

\iclrfinalcopy 
\begin{document}

\maketitle
\begin{abstract}
While molecular pre-training has shown great potential in enhancing drug discovery, the lack of a solid physical interpretation in current methods raises concerns about whether the learned representation truly captures the underlying explanatory factors in observed data, ultimately resulting in limited generalization and robustness. Although denoising methods offer a physical interpretation, their accuracy is often compromised by ad-hoc noise design, leading to inaccurate learned force fields. To address this limitation, this paper proposes a new method for molecular pre-training, called sliced denoising (SliDe), which is based on the classical mechanical intramolecular potential theory. SliDe utilizes a novel noise strategy that perturbs bond lengths, angles, and torsion angles to achieve better sampling over conformations. Additionally, it introduces a random slicing approach that circumvents the computationally expensive calculation of the Jacobian matrix, which is otherwise essential for estimating the force field. By aligning with physical principles, SliDe shows a 42\% improvement in the accuracy of estimated force fields compared to current state-of-the-art denoising methods, and thus outperforms traditional baselines on various molecular property prediction tasks.

\end{abstract}
\section{Introduction}\label{sec:intro}
Molecular representation learning plays a crucial role in a variety of drug discovery tasks, including molecular property prediction~\citep{schutt2018schnet,schutt2021equivariant,tholke2021torchmdnet}, molecular generation~\citep{Generativemodels,TorsionalDiffusion}, and protein-ligand binding~\citep{Gao2022CoSPCP,OnionNet}. To overcome the challenge of insufficient labeled data, various molecular pre-training methods have been proposed to obtain a universal molecular representation, including the contrastive approach~\citep{fang2022molecular,wang2022molecular,stark20223d,li2022geomgcl} and the predictive approach~\citep{rong2020self,fang2021chemrl,zhu2022unified,liu23sde}. 

According to \cite{Bengio}, a good representation is often one that captures the posterior distribution of the underlying explanatory factors for the observed input data. Regarding molecular representation, we posit that an ideal representation must adhere to the underlying physical principles that can accurately and universally illustrate molecular patterns. However, the majority of existing pre-training methods draw inspiration from pre-training tasks in computer vision and natural language processing and thus overlook the underlying physical principles. 


Nevertheless, designing self-supervised tasks that align with physical principles remains challenging. To the best of our knowledge, only one kind of the unsupervised molecular pre-training method has an explicit physical interpretation, i.e.~the 3D denoising approach~\citep{SheheryarZaidi2022PretrainingVD,pmlr-v202-feng23c}, which aims to learn an approximate force field for molecules. However, we have found that this approximate force field largely deviates from the true force field, due to inappropriate assumptions such as assuming a molecular force field is isotropic in coordinate denoising~\citep{SheheryarZaidi2022PretrainingVD} or treating certain parts being isotropically in Fractional denoising~\citep{pmlr-v202-feng23c}. Consequently, existing denoising methods still harbor a significant bias from physical laws, which can hinder downstream results, as depicted by the experiments conducted in \cite{pmlr-v202-feng23c} and our own experiments in Appendix~\ref{section:app dft label}. Therefore, it remains an essential issue to design a denoising method that better aligns with physical principles.


It should be noted that energy function is a pivotal factor in determining the quality of representation learning in denoising methods. Firstly, the Boltzmann distribution used for noise sampling, which determines the conformations on which the network learns its force field, is derived from the energy function. Secondly, the learned force field, which aims to align regression targets with the true molecular force field, is designated by the gradient of the energy function. As a result, a precise energy function facilitates the network to acquire accurate force fields for typical molecules, consequently enhancing the physical consistency of the representation.

Following the aforementioned analysis, we suggest utilizing the classical mechanical intramolecular potential energy function and approximating it in the quadratic form using relative coordinates, i.e.~bond lengths, angles, and torsion angles, with certain parameters. Inspired by the previous theoretical findings that associate the quadratic energy function with a Gaussian distribution through the Boltzmann distribution, we then propose a novel noise strategy, called BAT noise. Specifically, BAT noise introduces Gaussian noise to bond lengths, angles, and torsion angles, and their respective variances are predetermined by parameters within the energy function. This approach allows BAT noise to better approximate the true molecular distribution when compared to other existing methods. The resulting conformations from our strategy are closer to common low-energy structures than previous approaches, providing an advantage for effective representation learning.

The objective of the denoising target is to regress the molecular force field, i.e.~the gradient of the energy function w.r.t. Cartesian coordinates. However, the energy function is defined in relative coordinates, thus requiring a change of variables in the differential. Specifically, the gradient of the energy function in relation to relative coordinates is readily acquirable in the form of the by-term product of the BAT noise and the parameter.  Applying a variable change requires estimation of the Jacobian matrix of the coordinate transformation function, which is nevertheless computationally expensive. To address this issue, we introduce a random slicing technique that converts the Jacobian estimation into simple operations of coordinate noise additions and BAT noise acquisitions. 

Thus we have developed a novel and efficient method, known as sliced denoising (SliDe), which is equivalent to learning the force field of the utilized energy function. Consequently, SliDe possesses the ability to align better with physical principles by estimating a more precise force field.
To facilitate the learning process, we introduce a Transformer-based network architecture that explicitly encodes relative coordinate information and generates equivariant atom-wise features tailored for the sliced denoising task. Our contributions are summarized as follows:


1) Methodologically, we suggest the use of physical consistency as a guiding principle for molecular representation learning, and under this principle, we develop a novel sliced denoising method and corresponding network architecture.

2) Theoretically, we derive BAT noise from the classical mechanical energy function and establish the equivalence between learning the force field and our sliced denoising method.

3) Experimentally, we demonstrate that SliDe outperforms existing pre-training methods in terms of physical consistency and downstream performance on QM9 and MD17 datasets. 

\section{Background}\label{sec:Background}
Denoising is a kind of self-supervised learning task in molecular representation learning and has achieved outstanding results in many downstream tasks \citep{zhou2023unimol,pmlr-v202-feng23c,SheheryarZaidi2022PretrainingVD,luo2022one,ShengchaoLiu2022MolecularGP,jiao2022energy}. It refers to corrupting original molecules by specific noise and training the neural networks to predict the noise, thus reconstructing the molecules. Significant benefit of denoising over other pre-training methods is that it has been proven to be equivalent to learning a molecular force field, which is physically interpretable.

Coordinate denoising (Coord) \citep{SheheryarZaidi2022PretrainingVD} involves the addition of Gaussian noise to atomic coordinates of equilibrium structures, with subsequent training of the model to predict the noise from the noisy input. They establish the equivalence between coordinate denoising and force field learning, under the assumption of isotropic Gaussian noise. For a given sampled molecule $\mathcal{M}$, perturb the equilibrium structure $\vx_0$ by $p (\vx|\vx_0)\sim \mathcal{N}(\vx_0,\tau_{c}^2 I_{3N})$, where $\vx$ denotes the noisy conformation, $N$ denotes the number of atoms in the molecule, and $I_{3N}$ is the identity matrix of size $3N$, the subscript $c$ stands for the coordinate denoising approach. Assume the molecular distribution satisfies the energy-based Boltzmann distribution w.r.t the energy function $E_{Coord}$, then
    \begin{align}
     \mathcal{L}_{Coord} (\mathcal{M}) &=E_{p (\vx|\vx_0)p (\vx_0)}||GNN_{\theta} (\vx) - (\vx-\vx_0)||^2\\
    &\simeq  E_{p (\vx)}||GNN_{\theta} (\vx) -(- \nabla _{\vx} E_{Coord}(\vx))||^2,
\end{align}
where $GNN_{\theta} (\vx)$ refers to a graph neural network with parameters $\theta$ that takes the conformation $\vx$ as input and returns node-level predictions. The notation $\simeq $ indicates the equivalence between different optimization objectives for the GNN. The proof is supplemented in the appendix~\ref{sec:app proof}.

To account for the anisotropic molecular distribution, fractional denoising (Frad) \citep{pmlr-v202-feng23c} proposes introducing a hybrid noise on the dihedral angles of rotatable bonds and atomic coordinates and fractionally denoising the coordinate noise. This specially designed denoising task allows for a physical interpretation of learning force field. For a given sampled molecule $\mathcal{M}$, the equilibrium structure $\vx_0$ is perturbed by $p(\vpsi_a|\vpsi_0)\sim \mathcal{N}(\vpsi_0,\sigma_{f}^2  I_m)$ and $p(\vx|\vx_a)\sim \mathcal{N}(\vx_a,\tau_f^2  I_{3N})$, where $\vpsi_a$ and $\vpsi_0$ correpond to the dihedral angles of rotatable bonds in structures $\vx_a$ and $\vx_0$ respectively, with $m$ representing the number of rotatable bonds, and the subscript $f$ standing for Frad. Assume the molecular distribution satisfies the energy-based Boltzmann distribution w.r.t the energy function $E_{Frad}$, we have
\begin{align}\label{eq:Anisotropic denoising score matching}
      \mathcal{L}_{Frad} (\mathcal{M}) &= E_{p (\vx|\vx_a)p(\vx_a|\vx_0)p(\vx_0)}||GNN_{\theta} (\vx) - (\vx-\vx_a)||^2\\
    &\simeq  E_{p (\vx)}||GNN_{\theta} (\vx) -(- \nabla _{\vx} E_{Frad}(\vx))||^2.
\end{align}
 The proof is also supplemented in the appendix~\ref{sec:app proof}. A summary of denoising pre-training methods is provided in Appendix~\ref{sec:app Related Work Denoising}.

\begin{figure}[t]
    \begin{center}
    \centerline{\includegraphics[width=1\textwidth]{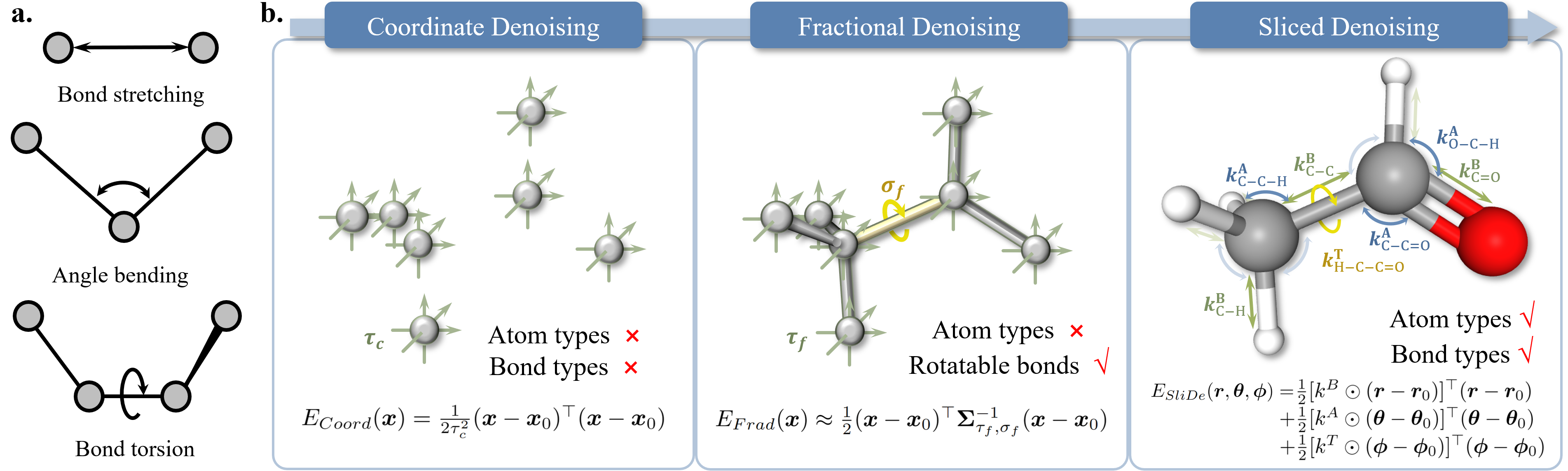}}
    \caption{\textbf{a}. Illustrations of bond stretching, bond angle bending, and bond torsion interactions. \textbf{b}. Comparison of the three denoising methods in terms of energy functions. Coordinate denoising learns an isotropic energy in Cartesian coordinates that does not discriminate different atom types and bond types. Based on coordinate denoising, fractional denoising treats the rotatable bonds in special. In contrast, sliced denoising performs fine-grained treatment for different atom types and bond types, enabling the most physically consistent description of the molecule.}
    \label{fig:illustrate}
    \end{center}
    \vskip -0.2in
\end{figure}

The aforementioned work has made efforts to learn physically interpretable molecular representations by designing noise distributions and their corresponding energy functions based on certain chemical priors. However, their energy function is coarse-grained as shown in Figure~\ref{fig:illustrate}b., lacking the capability to capture highly complex interaction information, such as bond stretching, angle bending, and bond torsion composed of different bond types and atom types. In contrast, our noise distributions and force fields are derived from a classical mechanical energy function, which is more consistent with the characteristics of true molecules. 
\section{Our Approach}
Inspired by the aforementioned deduction, we can conclude that in pursuit of designing an effective and interpretable denoising pre-training task, physical consistency can be achieved by developing an energy function that accurately approximates the true molecular energy. This, in turn, leads to a better noise distribution capable of sampling low-energy molecules, and a correspondingly improved force field learned through denoising. Following this guiding principle, we first establish a physical informed energy function in section~\ref{sec:energy function}, followed by the design of a noise distribution in section~\ref{sec:method,Noise design}. In section~\ref{sec:method,Sliced Denoising}, we present a denoising task aimed at learning the force field of the aforementioned energy function. Finally, in section~\ref{sec:method,Network Architecture}, we introduce the network architecture developed for our denoising method.

\subsection{Energy Function}\label{sec:energy function}
According to classical molecular potential energy theory ~\citep{Molecularsimu,MolecularDynamics}, the total molecular potential energy can be attributed to five types of interactions: bond stretching, bond angle bending, bond torsion, electrostatic, and van der Waals interactions. Figure~\ref{fig:illustrate}a. depicts the first three interactions. The energy function, in its general form, can be expressed as follows:
    \begin{equation}\label{eq:energy function complete}
\begin{aligned}
     E(  \vr,  \vtheta,  \vphi)&=\frac{1}{2}\sum_{i\in \mathbb{B}} k_i^{B} (r_i-r_{i,0})^2 +\frac{1}{2}\sum_{i\in \mathbb{A}} k_i^{A} (\theta_i-\theta_{i,0})^2\\
     &+\sum_{i\in \mathbb{T}} k_i^{T} (1-cos(\omega_i(\phi_i-\phi_{i,0})))
     + E_{elec} + E_{vdW},
\end{aligned}
\end{equation}
where $  \vr$, $  \vtheta$, and $  \vphi$ represent vectors of the bond lengths, bond angles, and bond torsion angles of the molecule, respectively. The index $i$ corresponds to the element in the vector. $  \vr_0$, $  \vtheta_0$, and $ \vphi_0$ correspond to the respective equilibrium values. The parameter vectors $  \vk^B$, $  \vk^A$, and $  \vk^T$ determine the interaction strength, while the parameter vectors $  \vomega$ determine the torsion periodicity. The index set $\mathbb{B}$, $\mathbb{A}$, $\mathbb{T}$ correspond to the bonds, angles, and torsion angles in the molecule, respectively. 

In order to approximate it as a quadratic form, which is often required to enable the equivalence based on previous proof, we put forward two mathematical operations. 
Firstly, when $\vphi \to   \vphi_0$, a Taylor expansion is utilized to express the bond torsion interaction in the quadratic form:
\begin{footnotesize}
    \begin{equation}
    1-cos(\omega_i(\phi_i-\phi_{i,0}))=1-[1-\frac{1}{2}(\omega_i(\phi_i-\phi_{i,0}))^2+o((\phi_i-\phi_{i,0})^2)]\approx\frac{1}{2}\omega_i^2(\phi_i-\phi_{i,0})^2.\notag
\end{equation}
\end{footnotesize}
The approximation is reasonable since the noise scale in denoising methods is usually small.  
Secondly, the fourth and fifth terms are neglected since these long-range interactions account for only a small fraction of the total energy and are often diminished in previous molecular representation learning methods ~\citep{tholke2021torchmdnet, alon2021on}. 
Consequently, we can simplify the energy function into quadratic form:
    \begin{equation}\label{eq:energy function simplified}
\begin{aligned}
     E_{BAT}(  \vr,  \vtheta,  \vphi)=
     \frac{1}{2}\sum_{i\in \mathbb{B}} k_i^{B} (r_i-r_{i,0})^2 +
     \frac{1}{2}\sum_{i\in \mathbb{A}} k_i^{A} (\theta_i-\theta_{i,0})^2+
     \frac{1}{2}\sum_{i\in \mathbb{T}} k_i^{T}\omega_i^2(\phi_i-\phi_{i,0})^2.     
\end{aligned}
\end{equation}
In Figure \ref{fig:illustrate}b., we compare our energy function with that of Coord and Frad.
Their formulations provide a general outline of the energy function in an averaged manner using only one or two parameters. Unfortunately, they fail to capture the nuanced energetic characteristics of molecules. In contrast, our energy function carefully describes the impact of different atomic types and bond types on energy using specific parameters. Therefore, our approach is more closely aligned with the true physical properties of molecules.

\subsection{Noise Design}\label{sec:method,Noise design}

With the common assumption that the conformation distribution of a molecule follows the Boltzmann distribution~\citep{boltzmann1868studien}, i.e.~$p \propto exp(-E)$, we can derive the conformation distribution corresponding to our quadratic energy function. 
    \begin{gather}
     p(  \vr,  \vtheta,  \vphi) = \frac{1}{Z}exp(-E_{BAT}(  \vr,  \vtheta,  \vphi) ) \\
    = \prod_{i\in \mathbb{B}} \frac{1}{Z^B_i}exp(-k_i^{B} \frac{(r_i-r_{i,0})^2}{2})
    \prod_{i\in \mathbb{A}} \frac{1}{Z^A_i}exp(-k_i^{A} (\frac{\theta_i-\theta_{i,0})^2}{2})
    \prod_{i\in \mathbb{T}} \frac{1}{Z^T_i}exp(-k_i^{T} \omega_i^2\frac{(\phi_i-\phi_{i,0})^2}{2}), \label{eq:noise dis in Z}
\end{gather}
where $Z$, $Z^B_i$, $Z^A_i$, $Z^T_i$ are normalization factors. According to \eqref{eq:noise dis in Z}, the conformation distribution can be expressed as a joint distribution of independent Gaussian on bond lengths, bond angles, and torsion angles. Therefore we can outline the following noise strategy. 
\begin{definition}[BAT noise]
The BAT noise strategy refers to perturbing the equilibrium structure by adding independent Gaussian noise on every \textbf{b}ond length, \textbf{a}ngle and \textbf{t}orsion angle:
    \begin{equation}\label{eq:BATnoise}
    \vr\sim \mathcal{N}(\vr_0, diag(\frac{1}{\vk^B})), \vtheta\sim \mathcal{N}(\vtheta_0, diag(\frac{1}{\vk^A})),\vphi\sim \mathcal{N}(\vphi_0, diag(\frac{1}{\vk^T\odot\vomega^2})),
\end{equation}
where $\odot$ means multiply item by item, $diag(\cdot)$ represents a diagonal matrix whose diagonal elements are the elements of the vector. 
The variances are determined by the parameters that can be obtained in prior, such as from the parameter files of molecular simulation tools. 
\end{definition}
Detail implementations of the BAT noise can be found in Appendix~\ref{sec:app noise design}. Since $E_{BAT}$ approximates the true molecular energy function, the sampling distribution of BAT noise resembles the true molecular distribution. This guarantees realistic sampled conformations that are beneficial for learning effective representations.  

\subsection{Sliced Denoising}\label{sec:method,Sliced Denoising}
Since the energy function is based on bond lengths, angles, and torsion angles, the gradient of the energy function can be represented as a simple form with respect to the relative coordinate:
  \begin{equation}\label{eq:force field in BAT coordinate}
         \nabla_{\vd}E_{BAT}(  \vd) =[ \vk^B\odot(  \vr-  \vr_0),\vk^A\odot(  \vtheta-  \vtheta_0),\vk^T\odot\vomega^2\odot(  \vphi-  \vphi_0)]^\top,
    \end{equation}   
where $\vd=(  \vr,  \vtheta,  \vphi)$. However, we need to derive a simple expression for the gradient of our energy function with respect to Cartesian coordinates and ensure the learning of the force field of $E_{BAT}$ by minimizing
    \begin{equation}\label{eq:ff target}
    E_{p(\vx|\vx_0)}||GNN_\theta(\vx) -  \nabla_{\vx}E_{BAT}(  \vd(\vx))||^2.
\end{equation}
For this purpose, we propose expanding the gradient using the chain rule and expressing the force field target as the gradient of the energy function with respect to the relative coordinates and a Jacobian matrix of the coordinate transformation. A rigorous formulation is presented as follows.

Firstly, we define a coordinate transformation function for a molecule $\mathcal{M}$ that maps from Cartesian coordinates to relative coordinates:
    \begin{align}
     f^\mathcal{M}: \mathbb{R}^{3N}&\longrightarrow (\mathbb{R}_{\geq 0})^{m_1}\times([0,2\pi))^{m_2}\times([0,2\pi))^{m_3} \\
      \vx &\longmapsto    \vd=(  \vr,  \vtheta,  \vphi),  \notag
\end{align}
 where $m_1$, $m_2$, and $m_3$ are numbers of bonds, angles, and torsion angles respectively. The mapping is well-defined, as these values can be uniquely determined by the Cartesian coordinates. Although $\vtheta$, $\vphi$ are defined on a torus, we can establish a homeomorphism between the Euclidean space $\mathbb{R}^{m_2+m_3}$ and $([0,2\pi)\backslash \{p_i\})^{m_2}\times([0,2\pi)\backslash \{p_j\})^{m_3}$, where $p_i,p_j$ are any points in $[0,2\pi),i=1\cdots m_2, j=1\cdots m_3$~\citep{Zorich2016II}. As the denoising method only involves conformations in a small neighborhood $V$ around the equilibrium conformation $\vd_0$, we can select $p_i,p_j$ such that $V\in ([0,2\pi)\backslash \{p_i\})^{m_2}\times([0,2\pi)\backslash \{p_j\})^{m_3}$. Consequently, the coordinate transformation function defined on a neighborhood can be regarded as a mapping between the Euclidean spaces $\mathbb{R}^{3N} \longrightarrow \mathbb{R}^{M}$, where $M\triangleq m_1+m_2+m_3$.
 
Assume that the coordinate transformation function $f^\mathcal{M}$ is continuously differentiable with continuous partial derivatives. Then the force field can be expressed by 
    \begin{equation}
    \nabla_{\vx}E_{BAT}( f(\vx))^\top= \nabla_{\vd}E_{BAT}(\vd)^\top\cdot J(\vx),
\end{equation}
 where   
 \begin{footnotesize}$J(\vx)=
  \left(
 \begin{matrix}
   \frac{\partial f^\mathcal{M}_1(\vx)}{\partial x_1} &  \cdots &\frac{\partial f^\mathcal{M}_1(\vx)}{\partial x_{3N}} \\
   \vdots &  & \vdots\\
   \frac{\partial f^\mathcal{M}_M(\vx)}{\partial x_1} &  \cdots &\frac{\partial f^\mathcal{M}_M(\vx)}{\partial x_{3N}}  \\
  \end{matrix}
  \right) \in \mathbb{R}^{M\times 3N}$ \end{footnotesize}
  is the Jacobian matrix. Then the target in \eqref{eq:ff target} can be written as
    \begin{equation}\label{eq:cartesian force field loss}
     E_{p(\vx|\vx_0)}||GNN_\theta(\vx) -  \nabla_{\vd}E_{BAT}(\vd)^\top\cdot J(\vx)||^2.
\end{equation}
For each noisy conformation, the Jacobian matrix can be estimated via numerical differentiation, which is time-consuming. To efficiently learn the force field, we devise a cleverly designed asymptotically unbiased estimator that does not require the computation of the Jacobian matrix, by utilizing two computational techniques. 

Firstly, a random slicing technique is introduced to estimate the target regression loss through the projection of the GNN and force field onto random vectors, as illustrated by lemma \ref{lemma2}.

\begin{Lemma}[Random Slicing]\label{lemma2}
     $\forall \va,\vb,\vv\in \mathbb{R}^{3N}$, $\sigma>0$, $\vv\sim \mathcal{N}(\vzero,\sigma^2 I_{3N})$, then
    \begin{equation}
         ||\va-\vb||^2=\frac{1}{\sigma^2}E_\vv[(\va-\vb)^\top\cdot \vv]^2.
     \end{equation}
\end{Lemma}
An intuitive illustration of lemma \ref{lemma2} is that the L2 norm of the vector $\va-\vb$ equals the expectations of the vector projected onto Gaussian random vectors. It is important to note that our random slicing technique, which can be seen as a more general approach, has been inspired by the idea of Sliced Wasserstein distance\citep{WassersteinBarycenter}. In fact, the latter can be considered as a special case of our random slicing technique applied to the cost function of Wasserstein distance.

After that, the dot product of the Jacobian matrix and the random vector can be efficiently calculated with the assistance of lemma \ref{lemma1}. 
\begin{Lemma}[Differential of Coordinate Transformation Function]\label{lemma1}
    $\forall \vx, \vv\in \mathbb{R}^{3N}$ are a molecular conformation and Cartesian coordinate noise respectively, then
    \begin{equation}\label{eq:lemma1}
        J(\vx)\cdot \vv=f^\mathcal{M}(\vx+\vv)-f^\mathcal{M}(\vx)+\alpha(\vx;\vv),
    \end{equation}
    where $\alpha(\vx;\vv)=o(\vv)$ as $||\vv||\to 0$.
\end{Lemma}

Therefore, the ultimate loss function can be defined as follows.
    \begin{equation}\label{eq:SliDe loss}\small
    \begin{aligned}
       \mathcal{L}_{SliDe}    (\mathcal{M}) =E_{p(\vx|\vx_0)}\frac{1}{N_v}\sum_{i=1}^{N_v} \left[GNN_\theta(\vx)^\top\cdot \vv_i 
            -\frac{1}{\sigma} \nabla_{\vd}E_{BAT}(  \vd)^\top\cdot\left(f^\mathcal{M}(\vx+\sigma \vv_i)-f^\mathcal{M}(\vx)\right)\right]^2,
    \end{aligned}
    \end{equation}
where $\vv_i\sim \mathcal{N}(\vzero, I_{3N})$, $\sigma$ is a parameter. $\vx$ and $\vd$ are the Cartesian coordinates and relative coordinates of the structure after adding the BAT noise to the equilibrium structure, respectively. $ \nabla_{\vd}E_{BAT}(  \vd)$ is given in \eqref{eq:force field in BAT coordinate}, and $GNN_\theta(\vx)\in \mathbb{R}^{3N}$ denotes the prediction output of GNN for each atomic Cartesian coordinate.

Consequently, the total loss is averaged on every sample in the 3D equilibrium molecular dataset $\mathbb{M}$:
    \begin{equation}
    \mathcal{L}^{total}_{SliDe}=\frac{1}{|\mathbb{M}|}\sum_{\mathcal{M}\in\mathbb{M}}\mathcal{L}_{SliDe}(\mathcal{M}).
\end{equation}
The computation of the scalar target can be performed rapidly by leveraging the relative coordinate obtained after adding the Cartesian noise $v$ through the utilization of RDKit~\citep{landrum2013rdkit}, a readily available cheminformatics tool. For reference, the pseudo-code outlining the approach for performing SliDe denoising pre-training is presented in Appendix~\ref{app: Algorithms}. 

Furthermore, we have proven its equivalence to learning the force field of $E_{BAT}$ as shown in the following theorem, with the proof provided in appendix~\ref{sec:app proof}.
\begin{theorem}[Interpretation of Sliced Denoising]\label{thm:sliced denoise}
    Given equilibrium structures, when $\sigma$ approaches $0$ and $N_\vv$ approaches $\infty$, minimizing $\mathcal{L}_{SliDe}(\mathcal{M})$ is equivalent to learning the force field of $E_{BAT}$ in Cartesian coordinate in \eqref{eq:ff target}.     
\end{theorem}

   
\subsection{Network Architecture}\label{sec:method,Network Architecture}
Compared to previous denoising methods, our approach defines energy and noise w.r.t. relative coordinates. Relative coordinates provide a complete representation of molecular structure and conform the molecular symmetry, thereby offering advantages for molecular modeling. Further details about related work on 3D molecular modeling in relative coordinates can be found in Appendix~\ref{sec:app Related Work Relative 3D}. 

While TorchMD-NET~\citep{tholke2021torchmdnet} has achieved competitive results when applied in denoising tasks, as shown in ~\citep{SheheryarZaidi2022PretrainingVD,pmlr-v202-feng23c}, it employs Cartesian coordinates to inject geometry information and does not explicitly model the angles and torsion angles. Since our method explicitly utilizes relative coordinates to model energy and noise, we believe angular information is important for learning our force field target. Therefore, in addition to the vertex update in TorchMD-NET, we also incorporate edge update and introduce angular information in the edge embeddings. These edge embeddings are then utilized in the attention layer, which impacts the vertex update. Our network is denoted as the Geometric Equivariant Transformer (GET), and further details are outlined in Appendix \ref{sec:app architecture}.

\section{Experiments}
Our first experiment in section \ref{sec:ff accuracy} is concerned with whether our approach achieves better physical consistency, specifically in terms of force field accuracy, as compared to coordinate denoising and fractional denoising methods. Then in section \ref{sec:results-downstream}, we evaluate the performance of SliDe in comparison to state-of-the-art 3D pre-training methods on the benchmark datasets QM9 and MD17, in order to assess our model's ability for molecular property prediction. Furthermore, in section \ref{sec:results-ablation}, we conduct ablation studies concerning fine-tuning regularization and network architecture. Additional experiments related to hyperparameter analysis, the impact of physical consistency on downstream tasks, and implementation details can be found in the Appendix \ref{sec:app Supplementary experiments} and \ref{sec:app Detail Implementations}.


\subsection{Evaluations on Physical Consistency}\label{sec:ff accuracy}
To estimate the learned force field in SliDe, we calculate the Cartesian force field for each molecule $\mathcal{M}$ by solving a least square estimation problem $\mA\vx_f=\vb$, where $\mA=[v_1,\cdots,v_{N_v}]^\top \in \mathbb{R}^{N_v\times 3N}$,  $b_i= \frac{1}{\sigma} \nabla_{\vd}E_{BAT}(  \vd)^\top\left(f^\mathcal{M}(\vx+\sigma \vv_i)-f^\mathcal{M}(\vx)\right)$, $\forall i=1\cdots N_v$, $\vb=[b_1,\cdots,b_{N_v}]^\top\in \mathbb{R}^{N_v\times 3N}$. 
We can prove that the regression loss $E_{p(\mathcal{M}|\mathcal{M})}\left[GNN_\theta(\mathcal{M}) -\vx_f\right]^2$ is asymptotically an equivalent optimization problem to SliDe. Therefore $\vx_f$ can be viewed as the learned force field target in SliDe. Details can be found in appendix proposition \ref{prop:Slide regression loss}. 

To verify the accuracy of the learned force field in various denoising methods, we compare the correlation coefficient between the learned force field and the ground-truth force field calculated by density functional theory (DFT). Since obtaining the true force field label can be time-consuming, the experiment is carried out on $1000$ molecules randomly selected from dataset PCQM4Mv2 \citep{nakata2017pubchemqc}. The noisy conformations are generated according to each denoising method and the learned force fields of Frad and Coord are estimated as the approach in \cite{pmlr-v202-feng23c}. 
\begin{table}[htbp]  
    \caption{ Correlation coefficient between the learned force field and the ground-truth force field of the three methods. The standard deviation is shown in parentheses. The top results are in bold.}
    \label{tab:Force Field Accuracy}
    \begin{center}
    \begin{footnotesize}
    \begin{tabular}{lccc}
    \toprule
    	Denoising method & \makecell[c]{ Coord} &  \makecell[c]{ Frad} & \makecell[c]{ SliDe}      \\
        \midrule
        Correlation coefficient & 0.616(0.047) & 0.631 (0.046) & \textbf{0.895} (0.071)  \\        
    \bottomrule
    \end{tabular}
    \end{footnotesize}
    \end{center}
    \vskip -0.1in
\end{table}

The experimental results in Table \ref{tab:Force Field Accuracy} indicate that SliDe increases the correlation coefficient of the estimated force field by 42\%, compared to Frad and Coord. This confirms that the design of our energy function and sliced denoising can help the model learn a more accurate force field than other denoising methods, which is consistent with our theoretical analysis. In addition, our result on Frad and Coord is in line with the experimental results in \cite{pmlr-v202-feng23c}, although the experiments are carried out on different datasets. It has been verified in \cite{pmlr-v202-feng23c} that learning an accurate force field in denoising can improve downstream tasks. We also conduct a supplementary experiment in Appendix \ref{section:app dft label} to confirm the conclusion.
As a result, SliDe greatly improves the physical consistency of the denoising method, enabling the learned representations to have better performance on downstream tasks.

The validation of force field accuracy can also help us choose hyperparameters without training neural networks. Details are in Appendix \ref{sec:app analysis}.

\subsection{Evaluations on Downstream Tasks}\label{sec:results-downstream}
Our model is pre-trained on PCQM4Mv2 dataset \citep{nakata2017pubchemqc}, which contains 3.4 million organic molecules and provides one equilibrium conformation for each molecule. Following previous denoising methods, we apply the widely-used Noisy Nodes technique \citep{godwin2021simple}, which incorporates coordinate denoising as an auxiliary task in addition to the original property prediction objective in the fine-tuning phase. Nevertheless, we observe 
the hard optimization of Noisy Nodes in SliDe. To get the most out of the fine-tuning technique, we add a regularization term in pre-training loss when evaluating on QM9 and MD17 dataset, i.e.~$\left[GNN_\theta(\vx+\tau \vv)- \tau \vv\right]^2$. An ablation study on the regularization term is provided in section \ref{sec:results-ablation-regu}.
Hyperparameter settings for pre-training and finetuning are summarized in Appendix \ref{sec:app Hyperparameter}.

Our baselines include both 3D pre-training approaches, such as fractional denoising (Frad), coordinate denoising (Coord), 3D-EMGP~\citep{jiao2022energy}, SE(3)-DDM~\citep{ShengchaoLiu2022MolecularGP}, Transformer-M (TM)~\citep{luo2022one}, as well as supervised models such as TorchMD-NET (ET)~\citep{tholke2021torchmdnet}, SphereNet~\citep{Liu2022SphericalMP}, PaiNN~\citep{schutt2021equivariant}, E(n)-GNN\citep{satorras2021n},  DimeNet~\citep{gasteiger2020directional}, DimeNet++~\citep{klicpera2020fast}, SchNet~\citep{schutt2018schnet}. The results for these baselines are directly taken from the referred papers, except for Coord on MD17, which is produced by us due to its absence in their paper.
\begin{table}[t]
\setlength{\tabcolsep}{4pt}
    \caption{Performance (MAE $\downarrow$) on QM9. The best results are in bold. }
    \label{table:qm9}
    \begin{center}
    \begin{footnotesize}
    \scalebox{0.96}{
    \begin{tabular}{lcccccccccccc}
    \toprule
    	 &  \makecell[c]{$\mu$}	& 	 \makecell[c]{$\alpha$ }		&  \makecell[c]{homo }		& \makecell[c]{lumo}		& \makecell[c]{gap}	& \makecell[c]{$R^2$ }	& \makecell[c]{ZPVE	}	& \makecell[c]{$U_0$ }		& \makecell[c]{$U$ }		& \makecell[c]{$H$ }		& \makecell[c]{$G$} & \makecell[c]{$C_v$	}
     \\
     &  \scriptsize (D)	& 	 \scriptsize ($a_0^3$)		& \scriptsize  (meV)		&\scriptsize (meV)		&\scriptsize (meV)	& \scriptsize($a_0^2$)	&\scriptsize (meV)	&\tiny (meV)	&\tiny (meV)		&\tiny (meV)		&\tiny (meV) & \tiny\makecell[c]{($\frac{cal}{mol\cdot K}$)}\\
    \midrule
      SchNet & 	0.033 & 	0.235	 & 41.0 & 	34.0 & 63.0	 & 0.07	 & 1.70	 & 14.00	 & 19.00	 & 14.00	 & 14.00	 & 0.033\\
      E(n)-GNN & 	0.029	 & 0.071	 & 29.0	 & 25.0	 & 48.0	 & 0.11	 &   1.55	 & 11.00	 & 12.00	 & 12.00	 & 12.00	 & 0.031\\
      DimeNet++	 & 0.030	 & 0.044	 & 24.6	 & 19.5	 & 32.6	 & 0.33	 & 1.21	 & 6.32	 & 6.28 & 	6.53	 & 7.56	 & 0.023\\
      PaiNN	 & 0.012	 & 0.045	 & 27.6	 & 20.4	 & 45.7	 & 0.07	 & 1.28	 & 5.85	 & 5.83	 & 5.98	 & 7.35	 & 0.024\\
      SphereNet & 0.025 & 0.045 & 22.8 & 18.9 & 31.1  & 0.27 & \textbf{1.120} & 6.26&  6.36 & 6.33 &7.78 &0.022\\ 
    ET & 0.011 & 0.059 & 20.3 & 17.5 & 36.1 & \textbf{0.033} & 1.840 & 6.15 & 6.38 & 6.16 & 7.62 & 0.026 \\
   \midrule 
    TM &	0.037 &		0.041 &		17.5 &		16.2 &		27.4 &		0.075 &		1.18 &		9.37 &		9.41 &		9.39 &		9.63 &		0.022
     \\
       SE(3)-DDM 	 &	0.015	 &	0.046 &		23.5	 &	19.5	 &	40.2	 &	0.122	 &	1.31	 &	6.92 &		6.99	 &	7.09	 &	7.65	 &	0.024
     \\
      3D-EMGP &	0.020	 &	0.057	 &	21.3	 &	18.2	 &	37.1	 &	0.092	 &	1.38	 &	8.60 &		8.60	 &	8.70	 &	9.30	 &	0.026
    \\
    Coord	 &	0.012	 &	0.0517	 &	17.7	 &	14.3	 &	31.8	 &	0.4496	 &	1.71	 & 6.57  &		 6.11  &		 6.45  &		 6.91 
    	 &	0.020    \\  
    Frad &		0.010 &		0.0374	 &	15.3	 &	13.7	 &	27.8	 &	0.3419 &		1.418	 &	5.33 &	5.62	& 5.55  &	6.19
    	 &	0.020 \\ 
     SliDe &		\textbf{0.0087} &		\textbf{0.0366} &	\textbf{13.6}	 &	\textbf{12.3}	 &	\textbf{26.2}	 &	0.3405	 & 1.521	 &	\textbf{4.28}&\textbf{4.29}	& \textbf{4.26} &	\textbf{5.37}    	 &	\textbf{0.019} \\
    \bottomrule
    \end{tabular}
    }
    \end{footnotesize}
    \end{center}
    \vskip -0.2in
\end{table}
\subsubsection{QM9}\label{sec:results-qm9 }
QM9~\citep{ramakrishnan2014quantum} is a quantum chemistry dataset providing one equilibrium conformation and 12 labels of geometric, energetic, electronic, and thermodynamic properties for 134k stable small organic molecules made up of CHONF atoms. The data splitting follows standard settings which have a training set with 110,000 samples, a validation set with 10,000 samples, and a test set with the remaining 10,831 samples. The performance on 12 properties is measured by mean absolute error (MAE, lower is better) and the results are summarized in Table \ref{table:qm9}. 

First of all, our model achieves new state-of-the-art performance on 10 out of 12 tasks in QM9, reducing the mean absolute error (MAE) by 12.4\% compared to the existing state-of-the-art. Among them, SliDe performs particularly well on challenging energetic and thermodynamic tasks. We speculate that this is because these two tasks are more closely related to molecular potential energy and force fields that we focus on during pre-training, for instance, the potential energy is related to thermodynamic quantities as illustrated in~\citep{Saggion2019Thermodynamic}. It is worth noting that the downstream performance of the three interpretable methods, SliDe, Frad, and Coord, is in agreement with the result of learned force field accuracy in section~\ref{sec:ff accuracy}, i.e.~SliDe demonstrates the strongest performance while Frad outperforms Coord. These experimental findings once again confirm the importance of physical consistency to molecular representations.

\subsubsection{MD17}\label{sec:results-md17}
MD17~\citep{chmiela2017machine} is a dataset of molecular dynamics trajectories of 8 small organic molecules. 
For each molecule, 150k to nearly 1M conformations, corresponding total energy and force labels are provided. We choose the challenging force prediction as our downstream task. The data splitting follows a standard limited data setting, where the model is trained on only 1000 samples, from which 50 are used for validation and the remaining data is used for testing. The performance is also measured by mean absolute error and the results are summarized in Table \ref{table:md17}.

Despite the fact that the downstream task is closely related to our pre-training task, the input conformations in MD17 are far from equilibrium and the limited training data setting makes it even more challenging. In this case, we still achieve new state-of-the-art performance on the six molecules, indicating that the force field knowledge learned in SliDe pre-training is effectively transferred to the downstream force field task. 
\begin{table}[h]
\setlength{\tabcolsep}{4pt}
    \caption{Performance (MAE $\downarrow$) on MD17 force prediction (kcal/mol/ $\mathring{\textnormal{A}}$). The best results are in bold. *: PaiNN does not provide the result for Benzene, and SE(3)-DDM utilizes the dataset for Benzene from \cite{Chmiela_2018md}, which is a different version from ours~\citep{chmiela2017machine}.}
    \label{table:md17}
    \vskip 0.15in
    \begin{center}
    \begin{footnotesize}
    \begin{tabular}{lccccccccc}
    \toprule
      & Aspirin	 & 	Benzene	 & 	Ethanol	 & \makecell[c]{	Malonal\\-dehyde}	 & 	\makecell[c]{Naphtha\\-lene}	 & 	\makecell[c]{Salicy\\-lic Acid}		 & Toluene		 & Uracil	\\ 
        \midrule  	 
      SphereNet	 & 	0.430	 & 	0.178	 & 	0.208	 & 	0.340	 & 	0.178	 & 	0.360	 & 	0.155	 & 	0.267 \\ 
      SchNet	  & 	1.35	  & 	0.31	  & 	0.39	  & 	0.66	  & 	0.58		  & 0.85		  & 0.57		  & 0.56	\\ 
   DimeNet		  & 0.499	  & 	0.187	  & 	0.230	  & 	0.383	  & 	0.215		  & 0.374		  & 0.216	  & 	0.301	\\      
  PaiNN*		  & 0.338	  & 	-	  & 	0.224	  & 	0.319	  & 	0.077	  & 	0.195		  & 0.094	  & 	0.139	\\ 
     ET
	 & 0.2450	 & 	0.2187	 & 	0.1067	 & 	0.1667		 & 0.0593	 & 	0.1284		 & 0.0644	 & 	0.0887\\ 
    \midrule   
   SE(3)-DDM* 	 & 0.453	 & 	-	 & 	0.166	 & 	0.288	 & 	0.129	 & 	0.266	 & 	0.122	 & 	0.183\\ 
      Coord	 & 	0.2108	  &  	0.1692	  &  	0.0959	  &  \textbf{	0.1392}	  &  	0.0529	  &  	0.1087	  &  	0.0582	  &  	\textbf{0.0742}\\
   Frad 	 & 0.2087	 & 	0.1994	 & 	0.0910	 & 	0.1415	 & 	0.0530	 & 	0.1081	 & \textbf{	0.0540}	 & 	0.0760\\ 
   SliDe  &		\textbf{0.1740} &		\textbf{0.1691}	 &	\textbf{0.0882}	 &	0.1538	 &	\textbf{0.0483} &	\textbf{0.1006}	& \textbf{0.0540} &	0.0825\\
    \bottomrule
    \end{tabular}
    \end{footnotesize}
    \end{center}
    \vskip -0.2in
\end{table}
\subsection{Ablation study}\label{sec:results-ablation}
We conduct an ablation study to examine the impact of the regularization term introduced for better fine-tuning and to evaluate the performance of our modified network architectures.
\subsubsection{Regularization Term}\label{sec:results-ablation-regu}
To assess the effectiveness of the regularization term proposed for pre-training SliDe, we conduct pre-training with and without regularization and subsequently fine-tuned the models on three QM9 tasks. The network architecture remains consistent across all three setups, and the Noisy Nodes are implemented with the same configuration. The result is shown in the bottom three rows of Table \ref{table:regu ablation}. Our findings indicate that the regularization term can effectively improve the performance of downstream tasks. Notably, SliDe without regularization still outperforms training from scratch and yields similar performance to Frad. Moreover, we observe in experiment that the regularization reduces the Noisy Nodes loss downstream, suggesting that the regularization term contributes to optimizing Noisy Nodes. 
\vspace{-7pt}
\begin{table}[h]
\setlength{\tabcolsep}{4pt}
\begin{minipage}[t]{0.5\textwidth}
    \caption{Ablation study for regularization term. }
    \label{table:regu ablation}
    \begin{center}
    \begin{small}
    \begin{tabular}{lccc}
    \toprule
    	 QM9 & \makecell[c]{homo}		& \makecell[c]{lumo}		& \makecell[c]{gap}	\\
    \midrule
    Train from scratch	&17.6  &	16.7  &31.3\\
    SliDe w/o regularization &15.0&	14.8&	27.7\\ 
     SliDe w/ regularization &	\textbf{13.6}&	\textbf{12.3}&	\textbf{26.2}	 \\
    \bottomrule
    \end{tabular}
    \end{small}
    \end{center}
\end{minipage}
\begin{minipage}[t]{0.01\textwidth}
    \begin{tabular}{l}
    	 	\\
    \end{tabular}
\end{minipage}
\begin{minipage}[t]{0.49\textwidth}
\caption{Ablation study for network design.}
    \label{table:network ablation}
    \vskip 0.15in
    \begin{center}
    \begin{small}
    \begin{tabular}{lcccc}
    \toprule
    MD17 force prediction	 & \  Aspirin\ 	& \ 	Benzene \  \\
    \midrule    
    SliDe (ET) & 0.2045 & 0.1810 \\ 
     SliDe (GET) & \textbf{0.1740} & \textbf{0.1691}	 \\
    \bottomrule
    \end{tabular}
    \end{small}
    \end{center}
\end{minipage}
\end{table}
\vspace{-7pt}
\subsubsection{Network Design}\label{sec:results-ablation-network}
To show the advantage of the improved network to our SliDe, we pre-train the geometric equivariant Transformer (GET) and TorchMD-NET (ET) by sliced denoising and fine-tune them on MD17. As shown in Table \ref{table:network ablation}, our network further improves the performance, indicating the excellence of our novel network in depicting more intricate geometric features, such as angles and torsional angles.

\section{Conclusion}
This paper proposes a novel pre-training method, called sliced denoising, for molecular representation learning. Theoretically, it harbors a solid physical interpretation of learning force fields on molecular samples. The sampling distribution and regression targets are derived from classical mechanical molecular potential, ensuring more realistic input conformations and precise force field estimation than other denoising methods. Empirically, SliDe has shown significant improvements in force field estimation accuracy and various downstream tasks, including QM9 and MD17, as compared with previous supervised learning and pre-training methods. 

\bibliography{SLID}
\bibliographystyle{iclr2024_conference}

\appendix
\section{Proof of Theoretical Results}\label{sec:app proof}

\begin{proof}[Proof of Lemma \ref{lemma1}]
Since $f^\mathcal{M}$ is differentiable at
the point $\vx$, \eqref{eq:lemma1} is the definition of the differential of a function of several variables\citep{Zorich2015}.
\end{proof}
\begin{proof}[Proof of Lemma \ref{lemma2}]
    \begin{equation}\small
        \begin{aligned}
         E_\vv[(\va-\vb)^\top\cdot \vv]^2&=E_v[(\va-\vb)^\top \vv \vv^\top (\va-\vb) ]=(\va-\vb)^\top E_v[\vv \vv^\top] (\va-\vb)\\
         &=(\va-\vb)^\top \sigma^2 I_{3N} (\va-\vb)
         =\sigma^2||\va-\vb||^2   
        \end{aligned}
    \end{equation}
Divide both sides by $\sigma^2$, then the proof is completed.
\end{proof}
\begin{proof}[Proof of Theorem \ref{thm:sliced denoise}]
Let $\vv^{(\sigma)}\triangleq\sigma\vv$, $\vv\sim N(0, I_{3N})$.
    \begin{align}\small
     &\mathcal{L}_{SliDe}(\mathcal{M})\approx E_{p(\vx|\vx_0)}E_{\vv} \left[GNN_{\theta}(\vx)^\top\cdot \vv -\frac{1}{\sigma} \nabla_{\vd}E_{BAT}(  \vd)^\top\cdot\left(f^\mathcal{M}(\vx+\sigma \vv)-f^\mathcal{M}(\vx)\right)\right]^2 \label{eq:16} \\
     &=E_{p(\vx|\vx_0)}E_{\vv^{(\sigma)}} \frac{1}{\sigma^2}\left[GNN_{\theta}(\vx)^\top\cdot \vv^{(\sigma)} - \nabla_{\vd}E_{BAT}(  \vd)^\top\cdot\left(f^\mathcal{M}(\vx+\vv^{(\sigma)})-f^\mathcal{M}(\vx)\right)\right]^2\\
     &\approx E_{p(\vx|\vx_0)}E_{\vv^{(\sigma)}} \frac{1}{\sigma^2}\left[GNN_{\theta}(\vx)^\top\cdot \vv^{(\sigma)} - \nabla_{\vd}E_{BAT}(  \vd)^\top\cdot\left( J(\vx)\vv^{(\sigma)}\right)\right]^2\\
     &=E_{p(\vx|\vx_0)}E_{\vv^{(\sigma)}} \frac{1}{\sigma^2}\left[\left(GNN_{\theta}(\vx) -  J(\vx)^\top\cdot \nabla_{\vd}E_{BAT}(  \vd)\right)
    ^\top\cdot \vv^{(\sigma)})\right]^2\\
    &=E_{p(\vx|\vx_0)}||GNN_{\theta}(\vx) -  \nabla_{\vx}E_{BAT}(  \vd(\vx))||^2
    \end{align}
The first step of approximation holds in the sense that $\lim_{N_\vv\to \infty}\mathcal{L}_{SliDe}=\eqref{eq:16}$. The second step holds by substituting $\sigma\vv$ with $\vv^{(\sigma)}\sim N(0, \sigma^2 I_{3N})$. The third step holds because of Lemma~\ref{lemma1} and approximation holds in the sense that $\lim_{\sigma \to 0} \alpha(\vx;\vv^{(\sigma)})\to 0$. The fourth step of the equation is the associative and distributive law of vector multiplication. The last step uses Lemma~\ref{lemma2}.
\end{proof}
\begin{Prop}\label{prop:Slide regression loss}
     When $N_v\to\infty$ and the least square estimation referred in section \ref{sec:ff accuracy} gives the ground-truth force field, i.e. $\mA\vx_f=\vb$ for any sampled $\vv$, the following regression loss is an equivalent optimization problem to SliDe.
     \begin{equation}\label{eq:Slide regression loss}
    \begin{aligned}
    \mathcal{L}_{SliDe} ^{(reg)}   (\mathcal{M})      
        =E_{p(\mathcal{M}|\mathcal{M})}\left[GNN_{\theta}(\mathcal{M}) -\vx_f\right]^2,
    \end{aligned}
\end{equation}
\end{Prop}
\begin{proof}
When $N_v\to\infty$, by the law of large numbers, 
\begin{align}\small
    \lim_{N_v\to\infty}\mathcal{L}_{SliDe}   (\mathcal{M})
    &=\lim_{N_v\to\infty}E_{p(\vx|\vx_0)}\frac{1}{N_v}\sum_{i=1}^{N_v}  \left[\mA \cdot GNN_{\theta}(\vx)-\vb\right]^2\\
    &=E_{p(\vx|\vx_0)}E_\vv  \left[\mA\cdot GNN_{\theta}(\vx)-\vb\right]^2\triangleq \mathcal{L}_{SliDe} ^{(asymp)}   (\mathcal{M})
\end{align}
\begin{align}\small
    &\nabla \mathcal{L}_{SliDe} ^{(reg)}   (\mathcal{M}) = E_{p(\vx|\vx_0)} 2 \left(GNN_{\theta}(\vx) -\vx_f\right)^\top \nabla GNN_{\theta}(\vx)\\
    &\nabla \mathcal{L}_{SliDe}^{(asymp)}   (\mathcal{M}) = E_{p(\vx|\vx_0)}E_\vv  2 \left(\mA \cdot GNN_{\theta}(\vx)-\vb\right)^\top \mA \nabla GNN_{\theta}(\vx)\label{eq:grad1}
\end{align}
By assumption, $\mA\vx_f=\vb$ for any sampled $\vv$, then \eqref{eq:grad1}
   \begin{align}\small
   &= E_{p(\vx|\vx_0)}E_\vv 2 \left( GNN_{\theta}(\vx)-\vx_f\right)^\top \mA^\top\mA \nabla GNN_\theta(\vx)\\
   &= E_{p(\vx|\vx_0)} 2 \left( GNN_{\theta}(\vx)-\vx_f\right)^\top E_\vv\left[\mA^\top\mA\right] \nabla GNN_{\theta}(\vx)\label{eq:grad2}
   \end{align}
Since $\vv_i\sim N(0, I_{3N})$, every element in $\mA=[v_1,\cdots,v_{N_v}]$ is i.i.d standard normal distribution. Therefore $E_\vv\left[\mA^\top\mA\right]= N_v\cdot  \mI_{3N}$, i.e. \eqref{eq:grad2}
\begin{align}\small
    &= E_{p(\vx|\vx_0)} 2 N_v \left( GNN_{\theta}(\vx)-\vx_f\right)^\top \nabla GNN_{\theta}(\vx)\\
    &=  N_v \nabla \mathcal{L}_{SliDe} ^{(reg)} (\mathcal{M})
\end{align}
Consequently, $\nabla \mathcal{L}_{SliDe}^{(asymp)}  =N_v \nabla \mathcal{L}_{SliDe} ^{(reg)}$, $N_v$ is a constant, then the two optimization target share the same minima.
\end{proof}
\begin{theorem}[Interpretation of Coordinate Denoising~\citep{SheheryarZaidi2022PretrainingVD}]
    Assume the conformation distribution is a mixture of Gaussian distribution centered at the equilibriums:
    \begin{equation} \small
    p(\vx) =\int p ( \vx | \vx_0)p(\vx_0),\ p (\vx|\vx_0)\sim \mathcal{N}(\vx_0,\tau_{c}^2 I_{3N})
    \end{equation}
    $\vx_0,\ \vx\in \mathbb{R}^{3N}$ are equilibrium conformations and noisy conformation respectively, $N$ is the number of atoms in the molecule. It relates to molecular energy by Boltzmann distribution $p(\vx) \propto exp(-E_{Coord}(\vx))$.
    
    Then given a sampled molecule $\mathcal{M}$, the coordinate denoising loss is an equivalent optimization target to force field regression:
\begin{align}\small
     \mathcal{L}_{Coord} (\mathcal{M})& =E_{p (\vx|\vx_0)p (\vx_0)}||GNN_{\theta} (\vx) - (\vx-\vx_0)||^2\label{eq:app coord loss} \\
    &\simeq  E_{p (\vx)}||GNN_{\theta} (\vx) -(- \nabla _{\vx} E_{Coord}(\vx))||^2,\label{eq:app coord target}
\end{align}
where $GNN_{\theta} (\vx)$ denotes a graph neural network with parameters $\theta$ which takes conformation $\vx$ as an input and returns node-level noise predictions, $\simeq $ denotes equivalent optimization objectives for GNN. 
\end{theorem}
\begin{proof}
 According to Boltzmann distribution, \eqref{eq:app coord target}$=E_{p (\vx)}||GNN_{\theta} (\vx) - \nabla _{\vx} \log p (\vx)||^2 $. By using a conditional score matching lemma~\citep{PascalVincent2011ACB}, the equation above $=E_{p (\vx|\vx_0)p(\vx_0)}||GNN_{\theta} (\vx) - \nabla _{\vx} \log p (\vx|\vx_0)||^2+T_1$, where $T_1$ is constant independent of $\theta$. Then with the Gaussian assumption, it becomes $E_{p (\vx|\vx_0)p(\vx_0)}||GNN_{\theta} (\vx) - \frac{\vx_0-\vx}{\tau_c^2}||^2+T_1 $. Finally, since coefficients $-\frac{1}{\tau^2}$ do not rely on the input $\vx$, it can be absorbed into $GNN_{\theta}$, thus obtaining \eqref{eq:app coord loss}.
\end{proof}

\begin{theorem}[Interpretation of Fractional Denoising~\citep{pmlr-v202-feng23c}]
 Assume the conformation distribution is a mixture distribution centered at the equilibriums:
    \begin{equation} \small
    p(\vx) =\int p (\vx|\vx_a)p(\vx_a|\vx_0)p(\vx_0), \ p(\vpsi_a|\vpsi_0)\sim \mathcal{N}(\vpsi_0,\sigma_{f}^2  I_m),\  p(\vx|\vx_a)\sim \mathcal{N}(\vx_a,\tau_f^2  I_{3N}),
    \end{equation}
    where $\vx_0,\ \vx_a,\ \vx\in \mathbb{R}^{3N}$ are equilibrium conformation and noisy conformations respectively, $\vpsi$ and $\vpsi_0$ are the dihedral angles of rotatable bonds in conformation $\vx$ and $\vx_0$, $m$ is the number of the rotatable bonds. It relates to molecular energy by Boltzmann distribution $p(\vx) \propto exp(-E_{Frad}(\vx))$.
    
    Then given a sampled molecule $\mathcal{M}$, the fractional denoising loss is an equivalent optimization target to force field regression:
\begin{align}\small
        \mathcal{L}_{Frad} (\mathcal{M}) &=E_{p (\vx|\vx_a)p(\vx_a|\vx_0)p(\vx_0)}||GNN_{\theta} (\vx) - (\vx-\vx_a)||^2 \label{eq:app frad loss}\\
    &\simeq  E_{p (\vx)}||GNN_{\theta} (\vx) -(- \nabla _{\vx} E_{Frad}(\vx))||^2, \label{eq:app frad target}
\end{align}
\end{theorem}
\begin{proof}
 According to Boltzmann distribution, \eqref{eq:app frad target}$=E_{p (\vx)}||GNN_{\theta} (\vx) - \nabla _{\vx} \log p (\vx)||^2$. By using a conditional score matching lemma~\citep{PascalVincent2011ACB}, the equation above $=E_{p (\vx,\vx_a)}||GNN_{\theta} (\vx) - \nabla _{\vx} \log p (\vx|\vx_a)||^2+T_2$, where $T_2$ is constant independent of $\theta$. Since the part in the expectation does not contain $\vx_0$, it equals to $E_{p (\vx,\vx_a,\vx_0)}||GNN_{\theta} (\vx) - \nabla _{\vx} \log p (\vx|\vx_a)||^2+T_2$. Finally, $-\frac{1}{\tau^2}$ can be absorbed into GNN and obtain~\eqref{eq:app frad loss}.
\end{proof}
\section{Supplementary experiments}\label{sec:app Supplementary experiments}
\subsection{Hyperparameter analysis}\label{sec:app analysis}
\begin{table}[htbp]  
    \caption{ Force field accuracy in different settings of $N_v$ and $\sigma$. The top results are in bold. }
    \label{tab:nv}
    \vskip 0.15in
    \begin{center}
    \begin{footnotesize}
    \begin{tabular}{lccccc}
    \toprule
    	& \makecell[c]{ $N_v=32$\\$\sigma=0.001$} &  \makecell[c]{ $N_v=64$\\$\sigma=0.001$} & \makecell[c]{ $N_v=128$\\$\sigma=0.001$} &  \makecell[c]{ $N_v=512$\\$\sigma=0.001$} &\makecell[c]{ $N_v=512$\\$\sigma=0.01$}     \\
        \midrule
        $\rho $ & 0.536(0.067) & 0.753(0.079) & 0.895 (0.071) &  \textbf{0.896} (0.067)& 0.893(0.072)\\
        MSE & 2.3e-4(1e-4)  & 1.5e-4(8e-5) &   \textbf{7.5e-5}(2e-4) &  7.6e-5(7e-5) & 0.53(0.20)\\    
        Scale & 0.73(0.13) &  \textbf{0.98} (0.14) & 1.05(0.14) &1.06(0.15)& 41.97(5.70)\\
    \bottomrule
    \end{tabular}
    \end{footnotesize}
    \end{center}
    \vskip -0.1in
\end{table}
 Since an accurate force field target contributes to learning effective representations, we can choose hyperparameters by utilizing the least square estimation of learned force field accuracy introduced in section~\ref{sec:ff accuracy}. This parameter selection strategy obviates the need for training neural networks, thereby rendering the process efficient and principled. Accordingly, we validated the accuracy of the learned force field for several combinations of hyperparameters $N_v$ and $\sigma$. The results are shown in Table \ref{tab:nv}. The accuracy is measured by Pearson correlation coefficient ($\rho$, the larger the better), mean squared error (MSE, the smaller the better), and "scale", which is the quotient of the mean absolute values between learned force fields and DFT force fields, and the best value is $1$. The value in the bracket is the standard deviation. 
 
 In Theorem \ref{thm:sliced denoise}, the best hyperparameter in theory is $N_v\to\infty$ and $\sigma\to 0$. However, a large sampling size $N_v$ leads to slow pre-training, and a small sampling standard deviation $\sigma$ results in higher numerical accuracy required. In experimental results, larger $N_v$ leads to better force field accuracy but the trend tends to saturate when $N_v>512$. This is mainly because in the pre-training dataset, the atomic numbers of the molecules are generally distributed between $20$ and $40$, i.e. $N_v=128>3N$ for most molecules and the least square error is small in this case.
 As for the standard deviation $\sigma$, it has a small impact on the correlation coefficient, but significantly affects the MSE and scale. After considering both accuracy and efficiency, we choose $N_v=128$ and $\sigma=0.001$.

\subsection{Effect of physical consistency on downstream tasks}\label{section:app dft label}

To verify whether learning an accurate force field in denoising can improve downstream tasks, we compare the existing denoising method with supervised pre-training on precise force field labels by DFT. Since DFT calculation can be time-consuming, we randomly select 10,000 molecules with fewer than 30 atoms from the PCQM4Mv2 dataset and calculate their precise force field label using DFT. The pre-trained model is fine-tuned on MD17 datasets, as shown in Table \ref{table:learn forcefield}. The results indicate that as the accuracy of the force field in pre-training tasks increases, the performance on downstream tasks improves. Note that there is a large gap from Frad to "DFT label supervised" compared to the improvement from training from Scratch to Coord and Coord to Frad, indicating that there is still a large room for downstream improvement along the idea of learning force fields. These findings motivate us to design a denoising pre-training task to learn accurate force fields.

\begin{table}[h!]
\setlength{\tabcolsep}{3pt}
    \caption{The performance (MAE) comparison between pre-training tasks with different force field accuracy.}
    \label{table:learn forcefield}
    \vskip 0.15in
    \begin{center}
    \begin{footnotesize}
    \begin{tabular}{lccccc}
    \toprule
    	 & \makecell[c]{Train from Scratch}	& \quad Coord \quad  &\quad Frad \quad		& DFT label supervised
     \\
    \midrule
 Aspirin (Force) & 0.253 & 0.250 & 0.248  & 0.236  \\ 
	
    \bottomrule
    \end{tabular}
    \end{footnotesize}
    \end{center}
    \vskip -0.1in
\end{table}

\section{Detail Implementations}\label{sec:app Detail Implementations}
\subsection{Noise Design}\label{sec:app noise design}
In Section \ref{sec:method,Noise design}, the BAT noise is independent noise on bonds, angles, and torsion angles, but in some cases, the independence is difficult to achieve. We make special treatments for these situations. Firstly, when an atom is connected to more than two atoms, the angles centered on this atom are dependent. In this case, we fix one edge and add noise to the angles involving this edge. Additionally, when there are ring structures in the molecule, the bond lengths, bond angles, and dihedral angles formed by the atoms in the ring are dependent. As a solution, We do not add noise to the bonds, angles, and torsion angles that are inside the ring. One reason is that RDkit does not support modifying bond lengths, bond angles, and dihedral angles inside rings. The other reason is that we attempt to add low-level independent Gaussian Cartesian coordinate noise to the atoms inside the ring to perturb them sufficiently. However, we find that its force field accuracy on certain molecules is much lower than without adding noise inside the ring. We speculate that this is because perturbing the atomic coordinates in the ring affects the surrounding angles and torsion angles.
The variance of the noise is determined by the parameters as shown in \eqref{eq:BATnoise}. Our parameters are obtained from the parameter files of Open Force Field v.2.0.0 (Sage)~\citep{sage}.
\subsection{Pseudocode for Algorithms}\label{app: Algorithms}
In this section, we present pseudocode to illustrate the pre-training algorithm of SliDe in Algorithm \ref{alg:SLIDE}.
\begin{algorithm}[htbp]
\caption{Sliced Denoising Pre-training Algorithm}\label{alg:SLIDE}
\begin{algorithmic}[1]
\Require
\Statex$GNN$: Graph Neural Network
\Statex$\mathbb{M}$: Unlabeled 3D molecular pre-training dataset
\Statex$T$: Training steps
\Statex$\sigma$: Standard deviation of the sampled coordinate noise
\Statex$N_v$: Sample number of coordinate noise
\Statex$\mathcal N$: Gaussian distribution
\While{$T \neq 0$}
    \State Randomly sample a molecule $\mathcal{M}$ with equilibrium structure $\vx_0$ from the dataset $\mathbb{M}$.
    \State Get the bond lengths, angles, torsion angles of $\mathcal{M}$, denoted as $(  \vr_0,  \vtheta_0,  \vphi_0)$. 
    \State Get the parameters of $\mathcal{M}$, denoted as $( \vk^B,\vk^A,\vk^T, \vomega)$. 
    \State Add BAT noise on the structure according to \eqref{eq:BATnoise} and get the perturbed structure $\vx$ and $\vd$
    \State Calculate $ \nabla_{\vd}E_{BAT}(  \vd)$ according to \eqref{eq:force field in BAT coordinate}.
    \For{i = $1,...,N_v$}
        \State Sample coordinate noise $\vv_i\sim \mathcal{N}(\vzero, I_{3N})$
        \State Calculate its corresponding relative coordinate changes $f^\mathcal{M}(\vx+\sigma \vv_i)-f^\mathcal{M}(\vx)$
        \State $Loss_i = \left[GNN_{\theta}(\vx)^\top\cdot \vv_i   -\frac{1}{\sigma} \nabla_{\vd}E_{BAT}(  \vd)^\top\cdot\left(f^\mathcal{M}(\vx+\sigma \vv_i)-f^\mathcal{M}(\vx)\right)\right]^2$
    \EndFor
    \State Optimise $Loss=\frac{1}{N_v}\sum_{i=1}^{N_v}Loss_i$ and update GNN
    \State $T = T - 1$

\EndWhile
\end{algorithmic}
\end{algorithm}

\subsection{Architecture Details}\label{sec:app architecture}
\begin{figure}[t]
\vskip 0.2in
\begin{center}
\centerline{\includegraphics[width=0.8\textwidth]{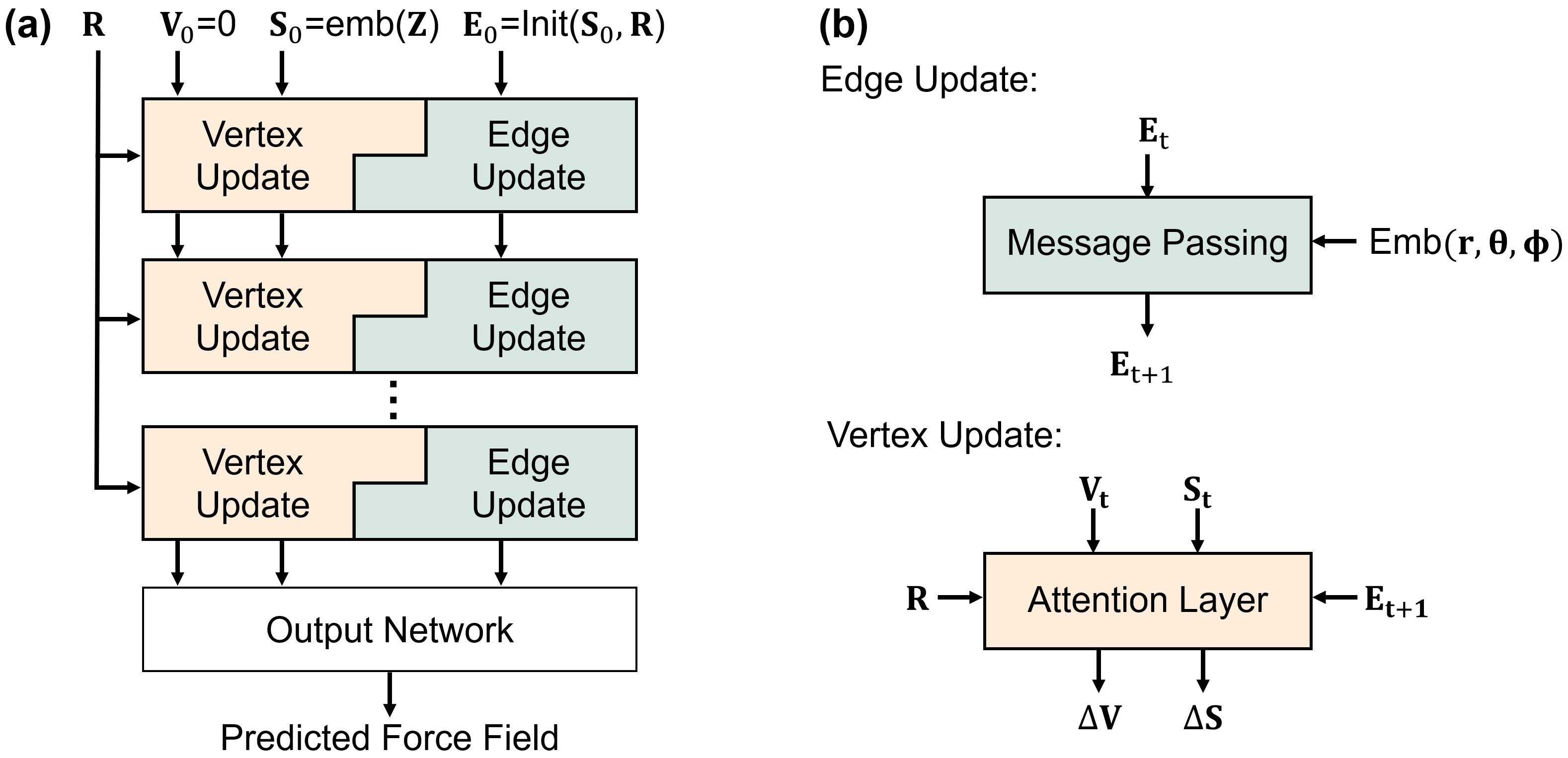}}
\caption{Overview of the network architecture. (a) The whole architecture includes initialization, several update layers and an output network. (b) The update layer consists of edge and vertex updates. The updated edge feature will be used in vertex updates.}
\label{fig:architectrue}
\end{center}
\vskip -0.2in
\end{figure}
Our network is an equivariant graph neural network that recursively updates vertex and edge features. An illustration of the network architecture is shown in Figure \ref{fig:architectrue}. The vertex feature $V\in \mathbb{R}^{3\times F_V}$ and $S\in \mathbb{R}^{F_V}$ are respectively vector and scalar features for each vertex, $F_V$ is the vertex feature dimension. $E\in \mathbb{R}^{F_E}$ denotes the edge feature of each edge, $F_E$ is the edge feature dimension. The edge vector $\vx_i-\vx_j$ is denoted by $R$. $Z$ is the atomic type. $\vr,\vtheta,\vphi$ are bond lengths, bond angles and torsion angles.

For the edge update, the invariant edge feature $E$ is updated by the embeddings of the bond length $\vr$, the neighbor edge features and the embeddings of their angles $\vtheta,\vphi$. Specifically, it incorporates Bessel functions to embed bond length and spherical harmonics to embed angle and torsion angles, which are shown to be effective geometry embeddings for molecular representation \citep{Liu2022SphericalMP,Klicpera2021GemNetUD,gasteiger2020directional}. For the vertex update, the invariant $S$ and equivariant $V$ are updated by an attention layer, whose architecture is based on TorchMD-NET. The updated edge features are projected into two filters and are later used to calculate attention weights in vertex update. 

\subsection{Hyperparameter Settings}\label{sec:app Hyperparameter}
\begin{table}[h!]
    \caption{Hyperparameters for pre-training. }
    \label{table:app setting pretrain}
    \vskip 0.15in
    \begin{center}
    \begin{footnotesize}
    \begin{tabular}{lc}
    \toprule
    Parameter & Value or description\\
     \midrule  
   Train Dataset & PCQM4MV2	\\
   Batch size & 	128	\\
    \midrule 
Optimizer  & 	AdamW	\\
Warm up steps & 	10000	\\
Max Learning rate & 	0.0004	\\
Learning rate decay policy	 & Cosine\\
	Learning rate factor & 	0.8\\
	Cosine cycle length	 & 240000\\
 \midrule 
Network structure	& \makecell[c]{Keep aligned with downstream settings \\respectively on QM9 and MD17}\\
 \midrule 
$N_v$ & 	128	\\
$\sigma$ & 	0.001	\\
Regression target & Least square results*\\
Regularization$^\dagger$ &  yes\\
$\tau$ & 0.04\\
   \bottomrule
    \end{tabular}
    \end{footnotesize}
    \end{center}
    \vskip -0.1in
\end{table}
Hyperparameters for pre-training are listed in Table \ref{table:app setting pretrain}. 
Details about Learning rate decay policy can be refered in \href{https://hasty.ai/docs/mp-wiki/scheduler/reducelronplateau#strong-reducelronplateau-explained-strong}{https://hasty.ai/docs/mp-wiki/scheduler/reducelronplateau\#strong-reducelronplateau-explained-strong}.

*: In precious denoising methods, normalizing the regression target, such as noise in Coord, is a widely applied technique to stabilize the training process. However, in SliDe loss \eqref{eq:SliDe loss}, the normalization is hard to implement. Instead, we chose to utilize $\mathcal{L}_{SliDe} ^{(reg)}$ and normalize the regression target $\vx_f$. We find the least squares estimation does not incur significant additional computational costs on the current dataset of molecular sizes.

$^\dagger$: In order to align with the downstream Noisy Node task that involves coordinate denoising, we add a regularization term according to \citep{pmlr-v202-feng23c} when evaluating on QM9 dataset. The loss is given by 
\begin{footnotesize}
    \begin{equation}
    \begin{aligned}
         &E_{p(\vx|\vx_0)}\frac{1}{N_v}\sum_{i=1}^{N_v}\left\{ \left[GNN_{\theta}(\vx+\tau \vv_i) - \tau \vv_i\right]^2 \right. \\ &+ \left.  \left[GNN_{\theta}(\vx)^\top\cdot \vv_i -\frac{1}{\sigma} \nabla_{\vd}E_{BAT}(  \vd)^\top\cdot\left(f^\mathcal{M}(\vx+\sigma \vv_i)- f^\mathcal{M}(\vx)\right)\right]^2 \right\}
    \end{aligned}
    \end{equation}
\end{footnotesize}

\begin{table}[h!]
    \caption{Hyperparameters for fine-tuning on MD17. }
    \label{table:app setting md17}
    \vskip 0.15in
    \begin{center}
    \begin{footnotesize}
    \begin{tabular}{lc}
    \toprule
    Parameter & Value or description\\
     \midrule  
  Train/Val/Test Splitting* &	950/50/remaining data	\\
  Batch size* &	8	\\
  \midrule  
Optimizer&	AdamW	\\
Warm up steps	&1000	\\
Max Learning rate	&0.001	\\
Learning rate decay policy&	ReduceLROnPlateau (Reduce Learning Rate on Plateau) scheduler	\\
	Learning rate factor&	0.8\\
	Patience	&30\\
	Min learning rate	&1.00E-07\\		
 \midrule  
Network structure	& Geometric Equivariant Transformer	\\
	Head number	&8	\\
	Layer number&	6	\\
	RBF number	&32	\\
	Activation function 	&SiLU	\\
	Embedding dimension&	128	\\
\midrule  
Force weight	&0.8		\\
Energy weight	&0.2		\\
Noisy Nodes(NN) denoise weight	&0.1		\\
Dihedral angle noise scale in NN&	20		\\
Coordinate noise scale in NN &	0.005		\\     
   \bottomrule
    \end{tabular}
    \end{footnotesize}
    \end{center}
    \vskip -0.1in
\end{table}
Hyperparameters for fine-tuning on MD17 are listed in Table \ref{table:app setting md17}. 

\begin{table}[h!]
    \caption{Hyperparameters for fine-tuning on QM9. }
    \label{table:app setting qm9}
    \vskip 0.15in
    \begin{center}
    \begin{footnotesize}
    \begin{tabular}{lc}
    \toprule
    Parameter & Value or description\\
     \midrule  
Train/Val/Test Splitting	 & 110000/10000/remaining data	\\
Batch size & 	128	\\
  \midrule  
Optimizer	 & AdamW	\\
Warm up steps & 	10000	\\
Max Learning rate	 & 0.0004	\\
Learning rate decay policy & 	Cosine	\\
	Learning rate factor	 & 0.8\\
	Cosine cycle length*	 & 300000 (500000) \\
  \midrule  		
Network structure & 	Geometric Equivariant Transformer	\\
	Head number	 & 8\\
	Layer number & 	8\\
	RBF number	 & 64\\
	Activation function  & 	SiLU\\
 Embedding dimension	 & 256\\
 Head&  \multirow{3}{*}{\makecell[c]{Applied according to \citep{tholke2021torchmdnet}}}\\ 
	Standardize	&\\
	AtomRef&\\
	  \midrule  
Label weight	 & 1	\\
Noisy Nodes denoise weight	 & 0.1(0.2)	\\
Coordinate noise scale	 & 0.005	\\   
   \bottomrule
    \end{tabular}
    \end{footnotesize}
    \end{center}
    \vskip -0.1in
\end{table}
Hyperparameters for fine-tuning on QM9 are listed in Table \ref{table:app setting qm9}. The cosine cycle length is set to be $500000$ for $\alpha$, $ZPVE$, $U_0$, $U$, $H$, $G$ and $300000$ for other tasks for fully converge.
 Following previous literature \citep{schutt2018schnet,schutt2021equivariant,Liu2022SphericalMP,ShengchaoLiu2022MolecularGP,pmlr-v202-feng23c}, we do not run cross-validation on QM9 and MD17, as the performance is quite stable for random seeds.
 
For experiment in section~\ref{sec:ff accuracy}, the force field calculated by DFT method is implemented by PySCF tool \citep{developmentsPySCF}, with basis = '6-31g', xc = 'b3lyp'.


Noisy Nodes is implemented following \citep{godwin2021simple,pmlr-v202-feng23c}.



\section{Related Work}\label{sec:app Related Work}
\subsection{Denoising for Molecular Pre-training}\label{sec:app Related Work Denoising}
Denoising as a self-supervised learning task originates from denoising generative models in computer vision \citep{PascalVincent2008ExtractingAC}. In molecular pre-training, it refers to corrupting and reconstructing the 3D structure of the molecule. Denoising is a self-supervised learning task designed specifically for 3D geometry data and achieves outstanding results in many downstream tasks for 3D molecules \citep{zhou2023unimol,pmlr-v202-feng23c}. 

The existing denoising methods mainly differ in the noise distribution and denoise tasks. Uni-Mol~\citep{zhou2023unimol} adds uniform noises of $[-1 \mathring{\textnormal{A}}, 1 \mathring{\textnormal{A}}]$ to the random $15\%$ atom coordinates. The model is trained to recover the correct atom coordinates and pair distance. They combine denoising with atom-type masking to make the masking task more challenging. 

Coordinate denoising (Coord) \citep{SheheryarZaidi2022PretrainingVD} adds Gaussian noise to atomic coordinates of equilibrium structures and trains the model to predict the noise from the noisy input. They establish the equivalence between coordinate denoising and force field learning.
Transformer-M~\citep{luo2022one} utilizes Coord to train the 3D model they proposed.

To capture the anisotropic molecular probability, fractional denoising (Frad) \citep{pmlr-v202-feng23c} proposes to add a hybrid noise on the dihedral angles of rotatable bonds and atomic coordinates, and fractionally denoise the coordinate noise. In this specially designed denoising task, the physical interpretation of learning force field also holds.

Compared to the aforementioned methods, our work most closely aligns with physical principles because our energy function better describes the true molecular energy landscape. This leads to a more realistic molecular force field and sampling distribution that is beneficial for representation learning.

On the other hand, to make the molecular energy invariant to rotation and translation, 3D-EMGP~\citep{ShengchaoLiu2022MolecularGP} denoises the Gaussian noise on the pairwise atomic distances and SE(3)-DDM~\citep{jiao2022energy} exploits the Riemann-Gaussian distribution for coordinate denoising. Our method naturally satisfies the symmetric prior because our energy function is defined on bond length, bond angle and dihedral angle, which are invariant to rotation and translation. 

\subsection{3D Molecular Modeling in Relative Coordinates}\label{sec:app Related Work Relative 3D}
 The geometric information contained in 3D conformers is crucial for molecular representation learning. Though most 3D structures are represented in Cartesian coordinates, recently many works have focused on utilizing 3D information in relative coordinates i.e. bond length, bond angle, torsion angle, also called as internal coordinates or local coordinates. The relative coordinates capture the complete geometry of atomic structures and are widely used because they are invariant to rotation and translation, making them convenient for molecular description in many scenarios~\citep{InternalCoordinatesProtein}. 

 For one thing, relative coordinates are used to enhance the expressiveness of graph neural networks. For molecular property prediction, SphereNet \citep{Liu2022SphericalMP} and GemNet \citep{Klicpera2021GemNetUD} encode bond length, bond angle and dihedral angle information by spherical Bessel functions and spherical harmonics functions. ALIGNN-d~\citep{Hsu2022EfficientGNNopticalspectroscopy} encode relative coordinates information by Radial Bessel basis and Gaussian basis and learn representations for optical spectroscopy prediction.
 
 For the other thing, the prediction of relative 3D information is found effective in pre-training task design. ChemRL-GEM\citep{fang2021chemrl} propose to predict bond lengths and bond angles to describe the local spatial structures. 3D PGT~\citep{Wang2023Automated3P} and GearNet~\citep{Zhang2022ProteinSR} also incorporate the prediction of bond length, bond angle and dihedral angle. They differ significantly from BAT-denoising in that their input structures remain unperturbed.
 
\end{document}